\documentclass[reprint,aps,onecolumn,11pt,notitlepage]{revtex4-1}
\usepackage[english]{babel}
\usepackage{amsmath,amssymb,graphicx,bm}
\usepackage[colorlinks=true, linkcolor=red, citecolor=blue]{hyperref}
\usepackage[justification=raggedright,font=small,labelfont=bf]{caption}
\usepackage{mathtools}
\usepackage{subfig}
\usepackage{float}
\usepackage[table,xcdraw]{xcolor}
\usepackage{booktabs}

\begin{document}
\title{Thermodynamic modifications to Bardeen black holes surrounded by quintessence based on the new higher order GUP}

\author{Bo-Li Liu $^{a}$}

\author{Yu Zhang $^{a}$ }
\email{zhangyu$_$128@126.com}
\author{Qi-Quan Li $^{b}$}
\author{Chen-Hao Xie $^{a}$}

\affiliation{
  $^a$Faculty of Science, Kunming University of Science and Technology,  Kunming, Yunnan 650500, China \\ $^b$School of Physics Science and Technology, Xinjiang University, Urumqi 830046, China\\}

\begin{abstract}
In this article, the thermodynamic properties of Bardeen black holes surrounded by quintessence are investigated in the framework of a new higher order GUP. The modified Hawking temperature, entropy and heat capacity are derived using a heuristic approach. Meanwhile, the remnant temperature and mass are deduced, and the modified black hole state equation is obtained by utilizing the energy density of matter. Ultimately, we analyze the effects of the GUP controlling deformation parameter $\beta $ on these thermodynamic properties using graphical illustrations, to more comprehensively understand the thermodynamic behavior of the black hole in the context of higher order GUP.
\end{abstract}
\maketitle

\section{Introduction}
In the field of physics, black holes have always been an enduring research topic. Because the laws of physics break down near a singularity and general relativity does not apply, scientists began looking for black holes without singularities. Firstly, Bardeen \cite{bardeen:1968non} proposed a regular black hole without singularities. Subsequently, researchers' discussions about other regular (non-singular) black holes have also been conducted continuously. Refs. \cite{Toshmatov:2014nya} and \cite{Ghosh:2014hea} respectively derived the exact solution for a rotating regular black hole and the solution for a radiative Kerr-like  regular black hole utilizing the Newman-Janis algorithm. Kumar et al. calculated the exact solution of the Bardeen black hole and the thermodynamic quantities related to the event horizon in the regularized four-dimensional Einstein-Gauss-Bonnet gravity framework \cite{Kumar:2020uyz}. As a thermodynamic system, regular black holes have always been the object of study for their entropy, heat capacity, Hawking temperature, and other thermodynamic properties. For example, Maluf and Neves discussed the thermodynamics of a class of regular black holes and pointed out the possible differences between regular and singular black holes \cite{Maluf:2018lyu}. Sharif and Saba Nawaz analyzed the thermodynamic properties of rotating Bardeen and Hayward regular black holes under both general conditions and Anti-de Sitter spacetimes, drawing conclusions related to thermodynamic stability \cite{Sharif:2020qgx}. According to relevant studies \cite{Jawad:2015qiu, Gonzalez-Diaz:2002por, Vagnozzi:2021quy,Escamilla:2023oce}, a large fraction of the energy density in the universe exists in the form of dark energy, which has a large negative pressure. To describe dark energy, researchers have proposed several dark energy models (e.g., the cosmological constant ), as well as scalar field models (e.g., quintessence). To probe the impact of dark energy on black holes, Ahmed Rizwan et al. studied the thermodynamics and geometric thermodynamics of regular Bardeen-AdS black holes with quintessence\cite{Rizwan:2018ozh}. Subsequently, Xie et al. investigated the chaotic behavior of strings surrounding Bardeen AdS black holes encompassed by typical dark energy \cite{Xie:2022yef}. Certainly, Kiselev \cite{Kiselev:2002dx} had proposed an exact spherical symmetric solution for a black hole surrounded by quintessence before this. L\"utf\"uo\u{g}lu et al. also studied the thermodynamic properties of the Schwarzschild black hole surrounded by quintessence under the GUP framework \cite{Lutfuoglu:2021ofc}.

For a long time, in order to improve the theory of quantum gravity, researchers have never stopped studying gravitational interactions. It should be pointed out that the development of quantum gravity is closely related to quantum mechanics. The Heisenberg Uncertainty Principle (HUP) cannot accurately measure the position and momentum of a particle at the same time, nor can it indicate the minimum observable length and maximum observable momentum. Some physical experiments have observed the existence of minimum lengths, such as Landau energy levels \cite{Das:2008kaa, Das:2009hs} and scanning tunneling microscopes \cite{Das:2009hs}. However, near the Planck scale, the HUP is no longer applicable and needs to be modified to the Generalised Uncertainty Principle (GUP), which is in favour of the existence of a minimum observable length \cite{Nozari:2012gd, Maggiore:1993kv, Ali:2009zq, Nozari:2011gj, Menculini:2013ida, Majumder:2011hy}. Moreover, string theory, loop quantum gravity, and black hole physics have also shown the need for corrections to the HUP \cite{Gross:1987kza, Amati:1988tn, Konishi:1989wk, Maggiore:1993rv, Bojowald:2011jd, Scardigli:1999jh, Adler:1999bu, Scardigli:2003kr}. GUP not only helps to understand the microstructure and thermodynamic properties of black holes, but may also provide explanations for some issues in classical black hole thermodynamics. For example, Banerjee and Ghosh proposed a new GUP that, when corrected for the thermodynamics of black holes, found that black hole evaporation stops at the stage of residual mass, and that this helped to avoid the singularity problem for black holes \cite{Banerjee:2010sd}. Bai et al. \cite{Bai:2022hti} utilized the GUP to correct temperature and phase transitions, and they employed thermodynamic geometry to determine the microscopic structure of black holes by taking the ratio of GUP parameters to charge. In addition, as GUP cannot be applied to large-scale length spaces, Park proposed the Extended Uncertainty Principle (EUP) \cite{Park:2007az}. And the linear combination of EUP and GUP can lead to the Extended Generalized Uncertainty Principle (EGUP) \cite{Bolen:2004sq}. The EGUP analysis was also utilized by Chen et al. \cite{Chen:2022dap} to calculate the thermodynamic quantities of RN black holes surrounded by quintessence . Meanwhile, in order to predict the upper limit of maximum momentum, Pedram \cite{Pedram:2011gw}proposed the higher order GUP, which implies the existence of minimum length uncertainty and maximum observable momentum, and is consistent with various quantum gravity theories. To obtain a unified expression for the thermodynamics of black holes, Hassanabadi et al. \cite{Hassanabadi:2019eol} used a simplified form of the higher order generalized uncertainty principle (GUP). It is worth noting that GUP may produce black hole remnants. Applying GUP to the study of black hole remnant states is of certain significance in helping to find potential candidates for dark matter \cite{Chen:2002tu, Li:2024tyk}. We learned that primordial regular black holes are possible candidates for dark matter in Refs. \cite{Calza:2024fzo,Calza:2024xdh}. In addition, the application of GUP in black hole thermodynamics also helps verify the validity of the second law of thermodynamics \cite{Chen:2017kpf, Okcu:2024tnw}.

The outline of this article is as follows. In Sec.~\ref{II}, we review the new higher order generalized uncertainty principle. Next, in Sec.~\ref{III}, we introduce the metric of a Bardeen black hole surrounded by quintessence , calculate its event horizon radius, and analyze the influence of related factors on the event horizon radius. Then, in Sec.~\ref{IV}, we derive the Hawking temperature, heat capacity and entropy functions as well as the state function equations under a novel higher-order GUP. Finally, Sec.~\ref{V} concludes with a summary of the results of the study.

\section{A New higher order generalized uncertainty principle} \label{II}
Das et al. \cite{Ali:2009zq, Das:2010zf, Ali:2011fa}, in order to introduce the maximum momentum, proposed a modified position-momentum commutation relationship,
\begin{equation}
           \left [ X_i, P_j \right ]= i\hbar\left [ \delta_{ij}-\alpha\left ( P\delta_{ij}+\frac{P_iP_j}{P} \right ) +\alpha^2 (P^2\delta_{ij}+3P_iP_j) \right ].
\end{equation}

Where $\alpha =\alpha _0/M_{pl}c=\alpha _0\ell_{pl}/\hbar $ is the GUP deformation parameter, $M_{pl}$ represents the Planck mass, $\ell _{pl}\approx 10^{-35}m$ denotes Planck length, and $M_{pl}c^2\approx 10^{19}GeV$  signifies the Planck energy.

However, there were some difficulties with the commutation relationship. Hence, Pedram suggested a higher order GUP based on this, which simultaneously includes minimum length uncertainty and maximum observable momentum \cite{Pedram:2011gw},
\begin{equation}\label{eq:E2}
\left [ X, P \right ] =\frac{i\hbar }{1-\beta P^2}.
\end{equation}

Among them, $\beta $ is a deformation parameter with quantum gravitational effects. It's worth mentioning that Eq.(\ref{eq:E2}) has a singularity at $P^2=1/\beta $, indicating the existence of an upper bound on momentum. The maximum observable value of momentum is $P=1/\sqrt{\beta}$, and the minimum length is $\frac{3\sqrt{3} }{4}\hbar\sqrt{\beta } $.

Hassanabadi et al. presented a simplified form of the higher order GUP \cite{Hassanabadi:2019eol},
\begin{equation}\label{eq:E3}
\left [ X, P \right ] =\frac{i\hbar }{1-\beta \left | P \right | }.
\end{equation}

The uncertainty relation corresponding to Eq.(\ref{eq:E3}) is
\begin{equation}
\bigtriangleup X\bigtriangleup P\ge \frac{\hbar }{2} \left [ -\beta(\bigtriangleup P)+\frac{1}{1-\beta(\bigtriangleup P)}   \right ].
\end{equation}

With the above uncertainty relation, we can get the uncertainty range of momentum as
\begin{equation}\label{eq:E5}
\frac{1}{2\beta }\left ( 1-\sqrt{1-\frac{4\hbar\beta  }{2(\bigtriangleup x)+\beta \hbar } }  \right )\le \bigtriangleup P\le \frac{1}{2\beta }\left ( 1+\sqrt{1-\frac{4\hbar\beta  }{2(\bigtriangleup x)+\beta \hbar } }  \right ).
\end{equation}

When the black hole captures particles, $ \bigtriangleup X$ should not be larger than the particular size corresponding to minimising the  horizon area. For a stationary spherically symmetric black hole, the size of the black hole is defined by the radius of the event horizon, and the range of positional uncertainty is given by the following formula:
\begin{equation}
2r_H\ge \bigtriangleup X\ge \frac{\hbar }{2} \left [ -\beta +\frac{1}{\bigtriangleup P(1-\beta \bigtriangleup P)}  \right ].
\end{equation}

 From the description, we learn about the form of the new higher order GUP. In the following sections, we will utilize this form to investigate the thermodynamic properties of Bardeen black holes surrounded by quintessence . The paper will uniformly use the natural unit system $c=G=\hbar=1$.

\section{Bardeen black hole surrounded by quintessence } \label{III}
The metric of a Bardeen black hole surrounded by the quintessence is \cite{Kiselev:2002dx,Merriam:2021bar,Lopez:2021ujg}
\begin{equation}
ds^2=-f(r)dt^2+\frac{1}{f(r)}dr^2 +r^2\left (d\theta ^2 +\sin^2 \theta +d\phi^2  \right ),
\end{equation}
where
\begin{equation}\label{eq:E6}
f(r)=1-\frac{2M}{r}+\frac{3Mg^2}{r^3}- \frac{\alpha }{r^{3w_q+1}}.
\end{equation}

Here, $ M $ represents the black hole mass, $ g $ denotes the magnetic charge, and $ \alpha  $ signifies a positive normalization factor that depends on the quintessence. The quintessence state parameter $w_q$, also known as the pressure index. To explain the accelerated expansion of the universe, physicists have inferred the existence of dark energy, which accounts for a high proportion of the total energy in the universe, about 68.3\%. Quintessence is a type of dark energy model, and the quintessence state parameter takes values in the range $-1 <  w_q <  -1/3  $.

When $f(r) = 0$, the radius of event horizon can be derived by solving the following equation:
\begin{equation}\label{eq:E7}
\left ( 1-\frac{2M}{r}+\frac{3Mg^2}{r^3}- \frac{\alpha }{r^{3w_q+1}} \right )\Bigg| _{r=r_H}  =0.
\end{equation}

In Eq.(\ref{eq:E6}) and Eq.(\ref{eq:E7}), we can see that the radius of the horizon is affected by the quintessence. Meanwhile, we can also gain the mass $M$ of the black hole,
\begin{equation}\label{eq:E10}
M=\frac{r_H^3}{2r_{H}^2-3g^2}\left (1-\frac{\alpha }{r_H^{3w_q+1}}   \right ).
\end{equation}

Among them, $r_H$ denotes the event horizon radius of the BH. Thereby, the surface gravity $\kappa $ of the black hole can be obtained,
\begin{equation}\label{eq:E11}
\kappa  =-\lim_{r \to r_H} \frac{\partial_rg_{tt}}{\sqrt{-g_{tt}g_{rr}} }= f(r)'=\frac{1}{r_H}\left ( 1-\frac{6g^2\left (1-\frac{\alpha }{r_H^{3w_q+1}}   \right ) }{2r_{H}^2-3g^2} +\frac{3w_q\alpha }{r_H^{3w_q+1}}  \right ).
\end{equation}
\begin{figure}
  \centering
  \includegraphics[width=0.5\linewidth]{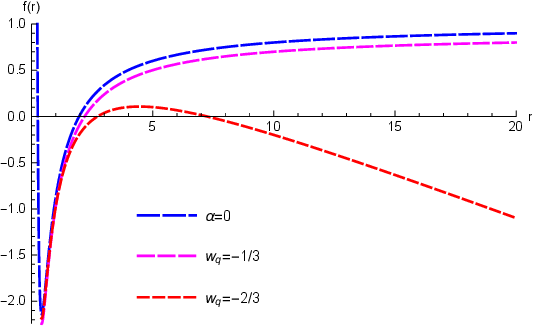}
  \caption{The behavior between $f (r)$ and $r$ when $\alpha=0,w_q=-1/3$ and $w_q=-2/3$.}
\label{liu26}
\end{figure}

To facilitate the discussion of the event horizon radius, we plot the relationship between $f(r)$ and $r$ in Fig.~\ref{liu26}. When $w_q = -1/3$, it can be seen that it aligns with the general case, presenting  one event horizon radius. At the same time, we observe that the inner event horizon and outer event horizon radii appear when $w_q = -2/3$.

For the radius of the event horizon, we will discuss two scenarios separately:
\begin{itemize}
\item{When  $w_q = -1/3$, three roots of the equation can be obtained, but only one has physical significance. The radius of  event horizon is as follows:}

\begin{equation}
    \begin{split}
        r_H & =\frac{2 M}{3 (1-\alpha )}+\frac{\left ( -81 \alpha ^2 g^2 M+162 \alpha  g^2 M-81 g^2 M+16 M^3+\chi  \right ) ^{1/3}}{3 \times 2^{1/3}  (1-\alpha )}\\&
        +\frac{4 \times 2^{1/3} M^2}{3 (1-\alpha ) \left ( -81 \alpha ^2 g^2 M+162 \alpha  g^2 M-81 g^2 M+16 M^3+\chi  \right ) ^{1/3}},
   \end{split}
\end{equation}
where
\begin{equation}
\chi =\sqrt{\left(-81 \alpha ^2 g^2 M+162 \alpha  g^2 M-81 g^2 M+16 M^3\right)^2-256 M^6}.
\end{equation}
\end{itemize}
\begin{itemize}
\item{When $w_q=-2/3$, the radius of the event horizon of a black hole can be expressed as}

\begin{equation}\label{eq:E14}
    \begin{split}
r_{H_1}&=\frac{1}{4 \alpha }- \frac{1}{2} \sqrt{\frac{1}{4 \alpha ^2}-\frac{4 M}{3 \alpha }-R_1-R_2}\\&
\pm\frac{1}{2} \sqrt{\frac{1}{2 \alpha ^2}-\frac{8 M}{3 \alpha }-\frac{\frac{1}{\alpha ^3}-\frac{8 M}{\alpha ^2}}{4 \sqrt{\frac{1}{4 \alpha ^2}-\frac{4 M}{3 \alpha }-R_1-R_2}}+R_1+R_2},
  \end{split}
\end{equation}

\begin{equation}\label{eq:E15}
    \begin{split}
r_{H_2}&=\frac{1}{4 \alpha }+\frac{1}{2} \sqrt{\frac{1}{4 \alpha ^2}-\frac{4 M}{3 \alpha }-R_1-R_2}\\&
\pm \frac{1}{2} \sqrt{\frac{1}{2 \alpha ^2}-\frac{8 M}{3 \alpha }+\frac{\frac{1}{\alpha ^3}-\frac{8 M}{\alpha ^2}}{4 \sqrt{\frac{1}{4 \alpha ^2}-\frac{4 M}{3 \alpha }-R_1-R_2}}+R_1+R_2},
  \end{split}
\end{equation}
with
\begin{equation}
R_0=81 g^2 M-16 M^3-432 \alpha  g^2 M^2,
\end{equation}
\begin{equation}
R_1=\frac{4\times  2^{1/3} \left(M^2-9 \alpha  g^2 M\right)}{3 \alpha  \left ( R_0+\sqrt{{R_0}^2-4 \left(4 M^2-36 \alpha  g^2 M\right)^3} \right ) ^{1/3} },
\end{equation}
\begin{equation}
R_2=\frac{\left ( R_0+\sqrt{{R_0}^2-4 \left(4 M^2-36 \alpha  g^2 M\right)^3} \right ) ^{1/3} }{3\times  2^{1/3}  \alpha }.
\end{equation}

We can see that there are two horizons in Fig.~\ref{liu26}. However, Eq.(\ref{eq:E14}) and Eq.(\ref{eq:E15}) are relatively complicated. The Cauchy horizon and event horizon need to be determined based on the specific conditions.
\end{itemize}

\section{Thermodynamic properties under the new higher order GUP} \label{IV}

In the semiclassical framework, assuming that the entropy $S$ of a black hole is no longer simply proportional to the area of the horizon area ($S = A/4\hbar$ ), but is a function of the area of horizon  $A$, the Hawking temperature of the black hole can be given as \cite{Xiang:2009yq}
\begin{equation}\label{eq:E17}
T=\frac{\kappa }{8\pi } \frac{\mathrm{d} A}{\mathrm{d} S}.
\end{equation}

When a particle is engulfed by a black hole, the particle's information may be unavailable to the outside world along with the fall into the black hole. According to information theory, the minimum unit of information loss is 1 bit, and the minimum change in entropy that corresponds to this is $\ln2$. The minimum change in area $\bigtriangleup A$ is proportional to the size and mass of the particle \cite{Bekenstein:1973ur}, as indicated by the following equation:
\begin{equation}\label{eq:E18}
\bigtriangleup A\sim Xm.
\end{equation}

In quantum mechanics, the paths of particle is indeterminate and can be described using wave packets. The width of the wave packet is expressed as the standard deviation of the distribution, i.e., the positional uncertainty. For the characteristic size of the particle, there is the relation: $X \sim \bigtriangleup X$.
Also the momentum uncertainty should not be greater than the mass $ (\bigtriangleup P\le m)$, considering the above relationship \cite{Xiang:2009yq}, Eq.(\ref{eq:E18}) can be rewritten as follows:
\begin{equation}
A\ge \bigtriangleup X\bigtriangleup P.
\end{equation}

The above facts show that when the black hole absorbs particle, the event horizon area of the black hole has the following relationship with entropy:
\begin{equation}
\frac{dA}{dS} \simeq \frac{\left ( \bigtriangleup A \right )_{min} }{\left ( \bigtriangleup S \right )_{min} }\simeq\frac{\gamma }{\ln2  }\bigtriangleup X\bigtriangleup P.
\end{equation}

Then, utilizing the heuristic method \cite{Maghsoodi:2019fca}, the Hawking temperature of the black hole can be expressed as
\begin{equation}\label{eq:E21}
T\simeq \frac{\kappa \gamma }{8 \pi \ln 2 } \bigtriangleup X\bigtriangleup P.
\end{equation}
Here, $ \gamma $ is the correction factor. When $\frac{\mathrm{d} A}{\mathrm{d} S} $ is substituted into Eq.(\ref{eq:E17}), the obtained result reproduces the semiclassical result. Simultaneously, the reference\cite{Hassanabadi:2021kuc} shows that the particle size is related to the position uncertainty, assuming the following relationship:
\begin{equation}\label{eq:E22}
\bigtriangleup X\simeq 2r_H
\end{equation}

By linking Eq.(\ref{eq:E22}) and substituting Eq.(\ref{eq:E11}) and Eq.(\ref{eq:E5}) into Eq.(\ref{eq:E21}), we can obtain the relationship between the temperature of a black hole and the radius of its event horizon, denoted as
\begin{equation}
T=\frac{\gamma }{8 \pi  \beta  \ln2  }  \left(1-\frac{6 g^2 \left(1-\frac{\alpha }{r_H^{3 w_q+1}}\right)}{2 r_H^2-3 g^2}\right)\left(1-\sqrt{1-\frac{4 \beta }{\beta +4 r_H}}\right).
\end{equation}

When the parameter $\alpha  = \beta = g = 0$ in the above equation, i.e., without the inclusion of the quintessence as well as the magnetic charge and GUP deformation parameter, one has $T=\frac{\gamma }{16 \pi  \ln2  }$.
Therefore, to obtain the Hawking temperature $T=1/(4\pi r_H)$, $\gamma$ can be equal to $4 \ln2$ \cite{Ghaderi:2016dpi, Riess:1998dv, Carroll:1998zi, Armendariz-Picon:2000nqq, Caldwell:1999ew, Khoury:2003aq, Chen:2008ra, Wei:2011za, Gangopadhyay:2013ofa, Scardigli:2018jlm, Hassanabadi:2020osz, Buoninfante:2019fwr, Petruzziello:2020wkd}. The temperature of the Bardeen black hole surrounded by quintessence, after being corrected by the new higher order GUP, can be derived as
\begin{equation}\label{eq:E24}
T_{GUP}=\frac{1}{2 \pi  \beta }\left(1-\frac{6 g^2 \left(1-\frac{\alpha }{r_H^{3 w_q+1}}\right)}{2 r_H^2-3 g^2}+\frac{3 \alpha  w_q}{r_H^{3 w_q+1}}\right)\left(1-\sqrt{1-\frac{4 \beta }{\beta +4 r_H}}\right).
\end{equation}

When $\beta =0$, the above formula is reduced to the Hawking temperature of the Bardeen black hole surrounded by the quintessence under the HUP limit,
\begin{equation}
T_{HUP}=\frac{1}{4 \pi  r_H}\left ( 1-\frac{6 g^2 \left(1-\frac{\alpha }{r_H^{3 w_q+1}}\right)}{2 r_H^2-3 g^2}+\frac{3 \alpha  w_q}{r_H^{3 w_q+1}} \right ).
\end{equation}

While $\beta  = g = 0$ gives the Hawking temperature of Schwarzschild black hole surrounded by HUP-corrected quintessence,
\begin{equation}
T_{HUP}=\frac{1}{4 \pi  r_H}\left ( 1+\frac{3 \alpha  w_q}{r_H^{3 w_q+1}} \right ).
\end{equation}

Since the black hole temperature must be a positive and real value, it is necessary to make a constraint on the horizon radius. For the Bardeen black hole surrounded by quintessence, the first constraint condition is proposed in Ref. \cite{Wu:2022leh}, which is determined by the nature of the black hole. When $w_q = -1/3$, $r_H\ge  0. 4243$, and when $w_q = -2/3$, $0. 431 \le  r_H \le  4. 98$. According to Eq.(\ref{eq:E24}), the second constraint is determined by the GUP parameter, which requires the following restrictions on the horizon radius,
\begin{equation}
r_H\ge \frac{3\beta }{4}.
\end{equation}

\begin{figure}
  \centering
  \includegraphics[width=0.47\linewidth]{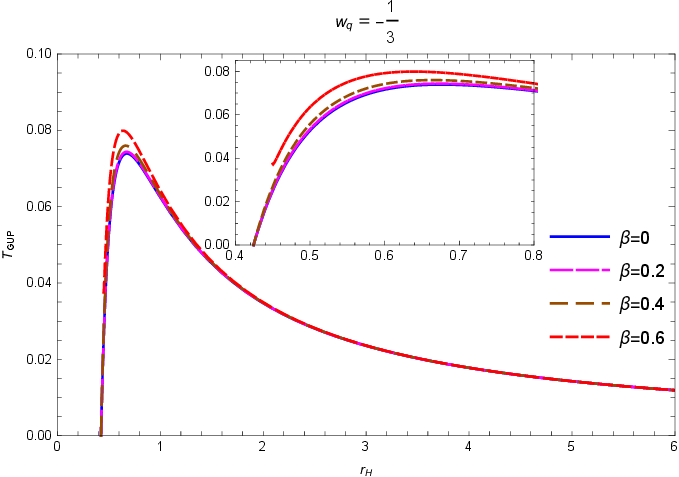}
  \includegraphics[width=0.47\linewidth]{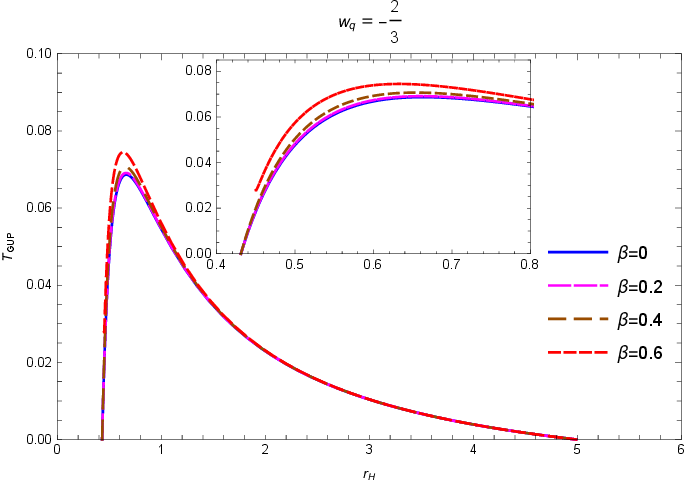}
  \caption{The behavior between $T_{GUP}$  and $r_H$ for different parameters $\beta$ $( \alpha =0.1, g=0.2)$.}
\label{liu6a7}
\end{figure}
For Fig.~\ref{liu6a7}, it can be seen that the effect of the GUP deformation parameter $ \beta$ on the Hawking temperature $T_{GUP}$ is mainly manifested in a section of the interval where the event horizon radius $r_H$ is small, implying that GUP has an impact in the early stage. The overall trend of the graph shows that the Hawking temperature first increases and then decreases for an increase in event horizon radius.

Next, we employ the thermodynamic relationship to derive the heat capacity function for the new higher order GUP correction,
\begin{equation}
C=\frac{d M}{d T},
\end{equation}
\begin{equation}\label{eq:E29}
C_{GUP}=\frac{3 \alpha  g^2 \left(2-3 w_q\right) r_H^{1-3 w_q}-9 g^2 r_H^2+6 \alpha  w_q r_H^{3-3 w_q}+2 r_H^4}{\left(3 g^2-2 r_H^2\right){}^2 \left(\frac{3 \varepsilon  \left(\sqrt{1-\frac{4 \beta }{\beta +4 r_H}}-1\right) r_H^{-3 w_q-2}}{2 \pi  \beta  \left(3 g^2-2 r_H^2\right){}^2}-\frac{4 \left(\frac{6 g^2 \left(1-\alpha  r_H^{-3 w_q-1}\right)}{3 g^2-2 r_H^2}+3 \alpha  w_q r_H^{-3 w_q-1}+1\right)}{\pi  \left(\beta +4 r_H\right){}^2 \sqrt{1-\frac{4 \beta }{\beta +4 r_H}}}\right)},
\end{equation}
where
\begin{equation}
\varepsilon =3 \alpha  g^4 \left(3 w_q-2\right) \left(3 w_q+1\right)+12 \alpha  g^2 r_H^2 \left(1-3 w_q^2\right)-8 g^2 r_H^{3 w_q+3}+4 \alpha  r_H^4 w_q \left(3 w_q+1\right).
\end{equation}

It is worth noting that the black hole stops exchanging radiation with its surrounding space when the heat capacity is zero, which will lead to black hole remnant. When we take the horizon radius as
\begin{equation}\label{eq:E30}
r_{rem}=\frac{3 \beta }{4}.
\end{equation}

It is observable that Eq.(\ref{eq:E29}) approaches zero, confirming the existence of black hole remnants \cite{Lin:2022doa,Lin:2022eix}. Substituting Eq.(\ref{eq:E30}) into Eq.(\ref{eq:E24}) yields the remnant temperature as follows:
\begin{equation}
T_{rem}=\frac{1}{2 \pi  \beta }\left (1+\frac{16 g^2 \left(1-\alpha  \left(\frac{3}{4}\right)^{-3 w_q-1} \beta ^{-3 w_q-1}\right)}{8 g^2-3 \beta ^2}+3^{-3 w_q} 4^{3 w_q+1}\alpha   w_q \beta ^{-3 w_q-1}  \right ).
\end{equation}

Similarly, the remnant mass can be derived, as shown in the following equation:
\begin{equation}
M_{rem}=-\frac{9 \beta ^3 \left(1-\alpha  \left(\frac{3}{4}\right)^{-3 w_q-1} \beta ^{-3 w_q-1}\right)}{8 \left(8 g^2-3 \beta ^2\right)}.
\end{equation}

It should be pointed out that in the absence of quintessence and magnetic charge, the heat capacity function and the remnant temperature of the black hole, modified by the new higher order GUP, are respectively reduced to
\begin{equation}\label{eq:E33}
C_{GUP}=-\frac{1}{8} \pi  \left(\beta +4 r_H\right){}^2 \sqrt{1-\frac{4 \beta }{\beta +4 r_H}},
\end{equation}
\begin{equation}
T_{rem}=\frac{3\beta }{8}.
\end{equation}

The black hole becomes the classical Schwarzschild black hole when there is no quintessence and magnetic charge. For Eq.(\ref{eq:E33}), when $\beta =0$, the heat capacity reduces to the standard heat capacity $C=-2\pi  r^2$, but the remnant temperature value becomes zero. Thus the remnant temperature exists only if the deformation parameter $\beta $ is not equal to zero, indicating that GUP affects such thermodynamic properties.

Next, let's derive the expressions of the specific heat and remnant temperature of the new higher order GUP correction under certain state parameters:
\begin{itemize}
\item {For $w_q=-1/3$, the corrected specific heat, temperature and remnant mass of the black hole are given as}
\begin{equation}
C_{GUP}=\frac{\pi  \beta  \eta  \left(2 r_H^4-9 g^2 r_H^2\right) \left(\beta +4 r_H\right){}^2}{4 \left(-27 \beta  g^4-4 \beta  r_H^4+3 \beta ^2 (\eta +3) g^2 r_H+24 \beta  (\eta +2) g^2 r_H^2+48 (\eta -1) g^2 r_H^3\right)},
\end{equation}
where
$\eta =\sqrt{1-\frac{4 \beta }{\beta +4 r_H}}$,
\begin{equation}
T_{rem}=-\frac{3 (\alpha -1) \left(\beta ^2-8 g^2\right)}{2 \pi  \beta  \left(3 \beta ^2-8 g^2\right)},
\end{equation}
\begin{equation}
M_{rem}=\frac{9 (\alpha -1) \beta ^3}{8 \left(8 g^2-3 \beta ^2\right)}.
\end{equation}
\end{itemize}
\begin{itemize}
\item{ For $w_q = -2/3$, the above thermodynamic quantities are rewritten respectively as}
\begin{equation}
C_{GUP}=\frac{r_H^2 \left(2 r_H \left(6 \alpha  g^2-2 \alpha  r_H^2+r_H\right)-9 g^2\right)\left(3 g^2-2 r_H^2\right){}^{-2}}{ \frac{2 \left(\sqrt{1-\frac{4 \beta }{\beta +4 r_H}}-1\right) \left(9 \alpha  g^4-3 \alpha  g^2 r_H^2-6 g^2 r_H+2 \alpha  r_H^4\right)}{\pi  \beta  \left(3 g^2-2 r_H^2\right){}^2}-\frac{4 \left(-\frac{6 g^2 \left(\alpha  r_H-1\right)}{3 g^2-2 r_H^2}-2 \alpha  r_H+1\right)}{\pi  \left(\beta +4 r_H\right){}^2 \sqrt{1-\frac{4 \beta }{\beta +4 r_H}}}},
\end{equation}
\begin{equation}
T_{rem}=\frac{3 (3 \alpha \beta -2) \beta ^2+48 g^2 (1-\alpha  \beta )}{4 \pi  \beta  \left(8 g^2-3 \beta ^2\right)},
\end{equation}
\begin{equation}
M_{rem}=\frac{9 \beta ^3 (3 \alpha  \beta -4)}{32 \left(8 g^2-3 \beta ^2\right)}.
\end{equation}
\end{itemize}

\begin{figure}
  \centering
  \includegraphics[width=0.5\linewidth]{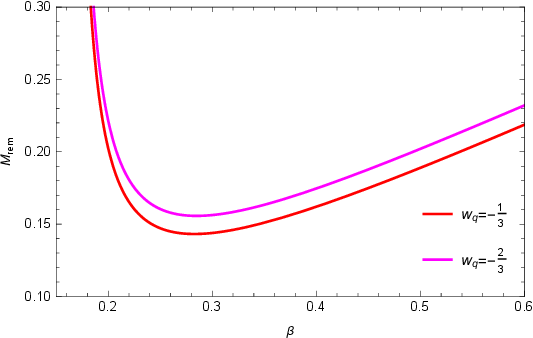}
  \caption{The behavior between $M_{rem}$ and $\beta$ for different parameters $w_q$.}
  \label{can1}
\end{figure}

Fig.~\ref{can1} shows that the corrected remnant mass $M_{rem}$ first decreases and then increases as the GUP deformation parameter $\beta$ increases.
\begin{figure}
  \centering
  \includegraphics[width=0.47\linewidth]{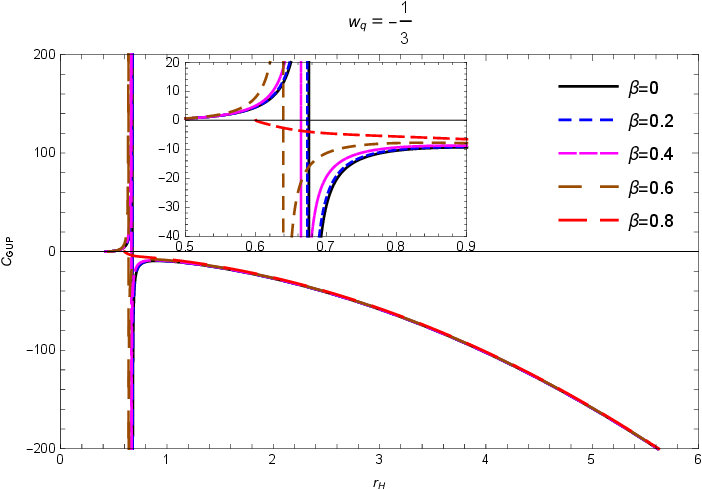}
  \includegraphics[width=0.47\linewidth]{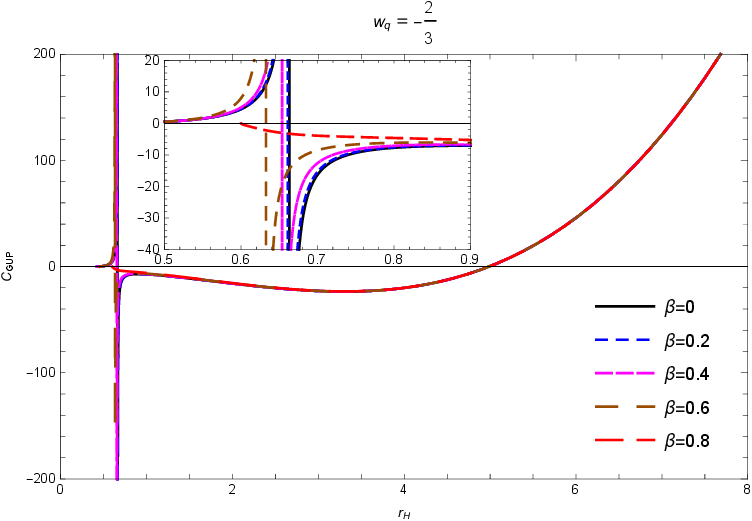}
  \caption{The behavior between $C_{GUP}$ and $r_H$ for different parameters $\beta $ $( \alpha =0.1, g=0.2)$.}
\label{liu8a9}
\end{figure}

It can be found that the GUP deformation parameter $\beta $ has a relatively weak effect on the overall trend of the black hole heat capacity $C_{GUP}$ in Fig.~\ref{liu8a9}. Nevertheless, closer inspection reveals that in regions with a small horizon radius, the graph has a tendency to move in the direction of decreasing horizon radius with increase in GUP deformation parameter, indicating that it has a certain effect on the black hole heat capacity as well as the black hole stability.

Then, we need to deduce the entropy within the framework of the new higher order GUP .
First of all, entropy is defined as
\begin{equation}\label{eq:E41}
S=\int \frac{d M}{T}.
\end{equation}

By substituting Eq.(\ref{eq:E10}) and Eq.(\ref{eq:E24}) into the above equation, we derive the entropy modified by the new higher order GUP,
\begin{equation}
\begin{split}
S_{GUP}&=\frac{1}{32} \pi  \left(-\beta ^2 \lambda _0+24 g^2 \left(\log \left(2 r_H^2-3 g^2\right)+2 \log \left(\beta  \left(\lambda _0-1\right)+4 \left(\lambda _0+1\right) r_H\right)\right)
\right. \\&
-2 \sqrt{6} g \lambda _1 \log \left(g \left(\beta  \left(\sqrt{6} \left(\lambda _1 \lambda _0-3 \beta \right)-12 g\right)  +4 r_H \left(\sqrt{6} \left(\lambda _1 \lambda _0-\beta \right)+12 g\right)\right)\right)\\&
+2 \sqrt{6} g \lambda _2 \log \left(g \left(\beta  \left(\sqrt{6} \left(\lambda _2 \lambda _0-3 \beta \right)+12 g\right)+4 r_H \left(\sqrt{6} \left(\lambda _2 \lambda _0-\beta \right)-12 g\right)\right)\right)\\&
-4 \sqrt{6} \beta  g \tanh ^{-1}\left(\frac{\sqrt{\frac{2}{3}} r_H}{g}\right)-2 \sqrt{6} g \lambda _2 \log \left(\sqrt{6} g+2 r_H\right)+2 \sqrt{6} g \lambda _1 \log \left(\sqrt{6} g-2 r_H\right)\\& \left.
-4 \beta ^2 \log \left(\beta  \left(\lambda _0-1\right)+4 \left(\lambda _0+1\right) r_H\right)+8 \beta  r_H+16 \left(\lambda _0+1\right) r_H^2\right),
\end{split}
\end{equation}
where
\begin{equation}
\lambda _0=\sqrt{1-\frac{4 \beta }{\beta +4 r_H}},
\end{equation}
\begin{equation}
\lambda _1=\sqrt{-3 \beta ^2+24 g^2-4 \sqrt{6} \beta  g},
\end{equation}
\begin{equation}
\lambda _2=\sqrt{-3 \beta ^2+24 g^2+4 \sqrt{6} \beta  g}.
\end{equation}

We can observe that the expression for the corrected entropy is intricate, prompting us to adopt a different perspective for discussion. First, let's revert to its calculation process.

The integrand of Eq.(\ref{eq:E41}) is
\begin{equation}\label{eq:E43}
\frac{d M}{T}=\frac{ \left(\pi  r_H^2 \left(\beta +4 r_H\right) \left(\sqrt{1-\frac{4 \beta }{\beta +4 r_H}}+1\right)\right)}{4 r_H^2-6 g^2}{dr}_H .
\end{equation}

It can be observed that the integral in Eq.(\ref{eq:E41}) is independent of the deformation parameter $\beta $. Eq.(\ref{eq:E41}) does not seem to include the quintessence state parameter $w_q$. However, we can observe that the horizon radius includes the quintessence state parameters by combining Eq.(\ref{eq:E6}) and Eq.(\ref{eq:E7}). Meanwhile, the entropy expression includes the event horizon radius.

When  $\beta= 0 $, Eq.(\ref{eq:E43}) becomes
\begin{equation}\label{eq:E44}
\frac{d M}{T}=\frac{4 \pi  r_H^3}{2 r_H^2-3 g^2}{dr}_H.
\end{equation}

Bringing Eq.(\ref{eq:E44}) back to Eq.(\ref{eq:E41}) and performing a Taylor expansion yields the general expression for entropy, $ S = \pi r^2 = A/4$. This calculation process shows that the quintessence does not directly affect the entropy of the Bardeen black hole. Instead, it influences the event horizon radius through the black hole metric, which in turn affects the entropy of black hole.

Finally, the pressure $P_q$ and the matter-energy density $\rho_q $ of a Bardeen black hole surrounded by quintessence are given by the following equation \cite{Kiselev:2002dx},
\begin{equation}\label{eq:E45}
\begin{split}
P_q&=\omega_q \rho_q , \\
\rho_q&=-\frac{3}{2} \frac{\alpha\omega_q  }{r^{3(\omega_q+1)}}.
\end{split}
\end{equation}

Substituting Eq.(\ref{eq:E45}) and Eq.(\ref{eq:E10}) into the volume $V=\frac{\partial M}{\partial P_q} $, we obtain \cite{Tharanath:2014uaa}
\begin{equation}\label{eq:E46}
V=\frac{r_H^3}{3w_q^2}.
\end{equation}

By relating Eq.(\ref{eq:E45}) and Eq.(\ref{eq:E46}) and substituting them into Eq.(\ref{eq:E24}), the Hawking temperature can be reformulated as
\begin{equation}
T_{GUP}=\frac{1}{2\pi \beta }\left ( 1-\frac{6g^2\left (1+2PV^\frac{2}{3}(3w_q^2)^{-\frac{1}{3}}    \right ) }{2(3w_q^2V)^\frac{2}{3}-3g^2} -\frac{2P(3Vw_q^2)^\frac{2}{3}}{w_q}  \right )\left (1- \sqrt{1-\frac{4\beta }{4(3w_q^2V)^\frac{1}{3}+\beta } }  \right ).
\end{equation}

Then for the $T = 1$ isotherm, we can obtain the form of the new higher-order GUP corrected black hole equation of state as
\begin{equation}\label{eq:E48}
P_{GUP}=\frac{2 w_q^2 \left(3 V w_q^2\right){}^{2/3}-9 g^2 w_q^2-\frac{1}{2} \pi  w_q^2 \left(2 \left(3 V w_q^2\right){}^{2/3}-3 g^2\right) \mu _1}{2 g^2 \left(2-3 w_q\right) \left(3 V w_q^2\right){}^{2/3}+4 w_q \left(3 V w_q^2\right){}^{4/3}},
\end{equation}
where
$$\mu _1=\left(1+\sqrt{1-\frac{4 \beta }{\beta +4 \sqrt[3]{3 V w_q^2}}}\right) \left(\beta +4 \sqrt[3]{3 V w_q^2}\right).$$

When $\beta $ is equal to zero, Eq.(\ref{eq:E48}) reduces to the equation of state of the black hole in the HUP limit:
\begin{equation}
P_{HUP}=\frac{3 g^2 \left(4 \sqrt[3]{3} \pi  \left (V w_q^2\right ) ^{1/3}-3\right) \left (V w_q^2\right ) ^{1/3}+2 V w_q^2 \left(3^{2/3}-12 \pi  \left (V w_q^2\right ) ^{1/3}\right)}{2 \sqrt[3]{3} V \left(\sqrt[3]{3} g^2 \left(2-3 w_q\right)+6 w_q \left(V w_q^2\right){}^{2/3}\right)}.
\end{equation}

To generate real-valued pressure in the context of higher order GUP, the following constraints are necessary:
\begin{equation}
V\geq \frac{1}{3 w_q^2}\left(\frac{3 \beta }{4}\right)^3.
\end{equation}

Based on the aforementioned facts, under specific values of quintessence state parameters, we have the following specific analysis:
\begin{itemize}
\item {When $w_q=- 1/3$,}
\begin{equation}
P_{GUP}=-\frac{g^2-\frac{2 V^{2/3}}{9\times 3^{2/3}}+\frac{1}{18} \pi  \left(\frac{2 V^{2/3}}{3^{2/3}}-3 g^2\right) \left(\beta +\frac{4 V^{1/3}}{3^{1/3}}\right) \left(1+\sqrt{1-\frac{4 \beta }{\beta +\frac{4V^{1/3}}{3^{1/3}}}}\right)}{2 \times 3^{1/3} g^2 V^{2/3}-\frac{4 V^{4/3}}{9 \times 3^{1/3}}}.
\end{equation}

In the HUP limit it will reduce to
\begin{equation}
P_{HUP}=\frac{3 g^2 \left(9-4\times 3^{2/3} \pi  V^{1/3}\right)-2\times  3^{1/3} V^{2/3}+8 \pi  V}{4\times  3^{2/3} V^{4/3}-54 \times 3^{1/3} g^2 V^{2/3}}.
\end{equation}
\end{itemize}
\begin{itemize}
\item {When $w_q=- 2/3$,}
\begin{equation}
P_{GUP}=\frac{162 g^2-24\times  6^{1/3} V^{2/3}+\pi  \left(4\times  6^{1/3} V^{2/3}-9 g^2\right)\mu _2 }{24 \left(2\times  6^{2/3} V^{4/3}-9\times  6^{1/3} g^2 V^{2/3}\right)},
\end{equation}
where
$$\mu _2=\left(3 \beta +4\times  6^{2/3} V^{1/3}\right) \left(1+\sqrt{1-\frac{12 \beta }{3 \beta +4\times 6^{2/3} V^{1/3}}}\right).$$
And with $\beta$ =0, the above equation transforms into
\begin{equation}
P_{HUP}=\frac{3 g^2 \left(9-4\times 6^{2/3} \pi  V^{1/3} \right)-4\times  6^{1/3} V^{2/3}+32 \pi  V}{8\times  6^{2/3} V^{4/3}-36\times  6^{1/3} g^2 V^{2/3}}.
\end{equation}
\end{itemize}
\begin{figure}
  \centering
  \includegraphics[width=0.47\linewidth]{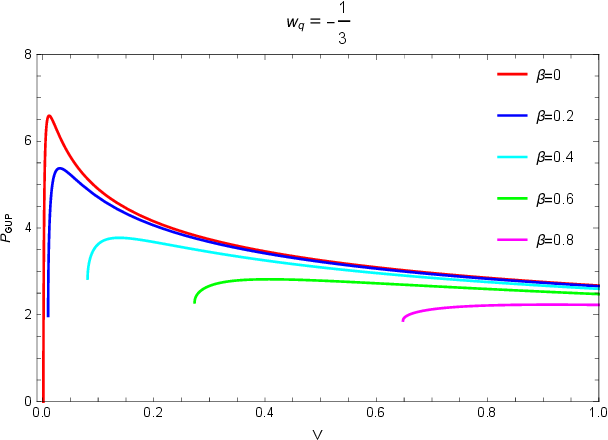}
  \includegraphics[width=0.47\linewidth]{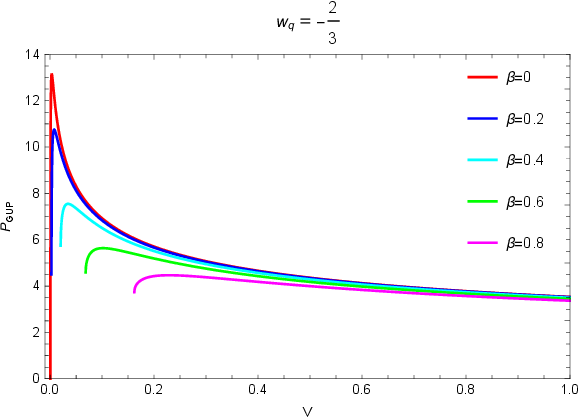}
  \caption{The behavior between $P_{GUP}$ and $r_H$ for different parameters $\beta $ $(g = 0. 002)$.}
\label{liu10a11}
\end{figure}

We plot the P-V isotherms for different cases in Fig.~\ref{liu10a11} to explore the effect of the higher order GUP deformation parameter $\beta$ versus the quintessence state parameter $w_q $ on the equation of state.
On the one hand, by comparing the left and right panels of Fig.~\ref{liu10a11}, we can see that the smaller the quintessence state parameter when the volume is relatively small, the more rapidly the pressure changes. On the other hand, the isotherm shifts in the direction of volume increase and the pressure decreases with the increase of the GUP deformation parameter.

\section{Conclusion} \label{V}
In this paper, we have investigated the impacts and corrections imparted by the new higher order GUP on the thermodynamics of Bardeen black holes surrounded by quintessence. To begin with, we calculated the event horizon radius of a Bardeen black hole surrounded by quintessence and discovered that quintessence influences the event horizon radius through the black hole metric. The second is the derivation of the corrected Hawking temperature using the heuristic method, and in order to satisfy the property that the corrected temperature has, Wu and his colleagues proposed a restriction on the horizon radius \cite{Wu:2022leh}. In this paper, another constraint on the horizon radius is calculated. Subsequently, we have investigated the corrected entropy, heat capacity, remnant temperature, remnant mass, as well as the state function equation of the black hole. The quintessence indirectly affects the entropy of the Bardeen black hole by affecting the event horizon radius. Furthermore, the impact of the GUP deformation parameter on the heat capacity of the black hole is mainly in the region with a small event horizon radius. For the corrected remnant mass, it exhibits a trend of first decreasing and then increasing as the deformation parameter increases, which suggests that the corrections imposed by the GUP do have an impact on the evaporation stage of black holes. Through the analysis of the P-V isotherm graphs after  the new higher order GUP corrections, it is found that both the GUP deformation parameter $\beta$  and the quintessence state parameter exert a certain influence on the isotherms.

\section*{Acknowledgments}
The work was supported by Yunnan Fundamental Research Projects (Grant No. 202301AS070029), and Yunnan Xingdian Talent Support Program - Young Talent Project.

\appendix

\end{document}